# Electromagnetic energy in a dispersive metamaterial

A. D. Boardman and K. Marinov[a]

Photonics and Nonlinear Science Group, Joule Laboratory, Department of Physics, University of Salford, Salford M5 4WT, United Kingdom

An expression for the electromagnetic field energy density in a dispersive, lossy, left-handed metamaterial, consisting of an array of split-ring resonators and an array of wires is derived. An electromagnetic field with general time-dependence is considered. The outcome is compared with previously published results. In the absence of losses, agreement with the general result for the energy density in a dispersive material is obtained. The formulae are verified using the finite-difference time-domain (FDTD) numerical method. The applicability of two commonly used permeability models to the problem of calculating the energy stored in an array of split-ring resonators is discussed.

1. **Introduction**

There is now a strong interest in the properties of the left-handed metamaterials (LHM) [1-3]. Because of this, formulating the electromagnetic field energy density in such materials has been addressed several times [4-7] and a number of different methods have been deployed.

As already pointed out [4] it ought to be the case that any conclusions that can be drawn concerning the electromagnetic energy density in a lossy dispersive material can be found easily in the literature. However, it appears that a precise answer is difficult to find and this is because there are no general formulations, valid for arbitrary materials. In the presence of dispersion and losses, the knowledge of the permittivity and permeability functions alone is insufficient to provide an expression for the stored electromagnetic energy density [7]. This is because a very detailed model of the microstructured medium under investigation is needed. Unfortunately, this means that the problem of finding the energy density has to be solved separately for every material.

---

[a] E-mail: k.marinov@salford.ac.uk

A long time ago, Loudon provided a beautiful discussion on how to include loss in the electromagnetic field energy [8]. In this age of metamaterials, it is important to see to what extent the arguments put forward by Loudon can still be used. In addition to the question of loss it is also important to demonstrate that even a metamaterial of the kind that is often called left-handed will still have a positive energy and avoid the possibility that a negative energy, which is unacceptable physically, might appear [4]. To address all these issues, a new discussion of the energy density in metamaterials is presented here. It not only makes contact with the original work of Loudon [8], but also with a recent and exciting work in the field [7].

Specifically, Loudon considered dielectrics with Lorentz-type of dispersion, and this has been generalized to include a material in which *both* the permittivity and the permeability are of Lorentz-type [5, 6]. The closed-form expressions that have emerged, coupled to the numerical calculations, show that the energy density is always causal and always positive.

Physically speaking, however, the arrays of split-ring resonators that provide the negative permeability in left-handed media (LHM) cannot be considered as a Lorentz-type of medium [9, 10]. To move the axiomatic Lorentz restriction a recent approach has produced a new expression for the energy density for such arrays [7]. This progress has been achieved, however, under conditions of time-harmonic excitation. The general case, using electromagnetic fields with arbitrary time-dependencies (e.g. short pulses) has not been considered yet; so this provision is one of the principle aims of the present study. The other one is to find a way of discussing the electromagnetic energy in LHM that is internally consistent, in the sense that it is robust with respect to low loss and high loss limits.

In this paper Loudon approach [8] permits the derivation of an expression for the energy density in an LHM consisting of a split-ring resonator array [9, 10] and an array of wires [3]. An arbitrary time-dependence of the electromagnetic field is assumed. The energy density is then compared to the, previously reported [7], time-harmonic electromagnetic field case and to the result for the energy density in Lorentz media [5-7]. It is shown that in a lossless, dispersive, material the result derived here reproduces the general formula for the electromagnetic energy density [11]. The new result for the energy density associated with the split-ring resonator array is used in conjunction with FDTD solutions of Maxwell's equations to show that energy conservation is satisfied to a high degree of accuracy.

## 2. Electromagnetic energy density in a left-handed metamaterial

The artificial molecules that make up a metamaterial, of the kind that have been labeled left-handed, are often composed of split-rings and metal wires. The latter provides the negative relative permittivity behavior while the former has the precise behavior of an equivalent LCR circuit under the restriction that the radius $r$ of the ring is much less than the electromagnetic wavelength $\lambda$. It is interesting that this is actually a very old problem and that the principal result, concerning the electromagnetic response of such an array, was published many years ago [9]. The recent popularity and applicability, however, has been driven by the work of Pendry.

Provided that the inequality $r \ll \lambda$ is substantially obeyed the treatment of the artificial molecule as an LCR circuit retains its validity. An actual metamaterial is a composite arrived at through a process of homogenization, however. To achieve this outcome, a split-ring array must be, initially, thought of as being part of particular lattice. Hence, if it is assumed that the rings are on a lattice with cubic symmetry the final outcome is an isotropic metamaterial. For simplicity, this is the symmetry assumed here, without loss of generality. A composite *isotropic* metamaterial, consisting of an array of split-ring resonators (SRR) and an array of wires, can be precisely investigated with an equivalent LCR circuit. The final outcome is a relative scalar permeability

$$\mu(\omega) = 1 + \frac{F\omega^2}{\omega_0^2 - \omega^2 - i\omega\gamma} \qquad (1)$$

and a relative scalar permittivity

$$\varepsilon(\omega) = 1 - \frac{\omega_p^2}{\omega(\omega + i\nu)}, \qquad (2)$$

[3, 9, 10], where $\omega$ is the excitation angular frequency, $\omega_p$ is the effective plasma frequency, $\omega_0$ is the resonant frequency and $\nu$ and $\gamma$ are the loss parameters. Note that (2) is used to model the behavior of a cold electron plasma. Equation (1) describes the response of an array of split-ring resonators to an external magnetic field [9, 10]. It is important to reemphasize that (1) can only be used provided that the ring radius is much less then the wavelength and this fits into the concept of a metamaterial as a composite of artificial subwavelength "molecules" [13]. The consequence of this assumption is that it permits the conduction current to dominate the displacement current. This physical situation is referred to in electromagnetics as a quasistatic approximation [11]. It is clear that (1) does not provide a description of

very high-frequency behavior because as $\omega \to \infty$ the permeability $\mu \to 1-F$, while it is $\mu \to 1$ that should be expected physically. There is straightforward physical way to see this by recognizing that the concept of an inductor and capacitor break down at very high frequencies. It does not mean, however, that the expression (1) is wrong just because of this limitation on its frequency behavior. Indeed, it has been suggested [14-18] that a Lorentz model,

$$\mu(\omega) = 1 + \frac{F\omega_0^2}{\omega_0^2 - \omega^2 - i\omega\gamma} \qquad (3)$$

is a suitable form for the relative permeability. Such a model, however, is axiomatic, since it is not proven by any microscopic considerations. At present the literature is populated by both models without apparent difficulty: some studies [14-18] use (3) while others [7, 19, 20] use (1). This is easily explained by the fact that the frequency range that is interesting for applications maps onto the resonance region $\omega \approx \omega_0$ and this is precisely where both models have similar behavior. Away from the resonance the model described in (1) fails as $\omega \to \infty$ because it leads to a relative permeability of $\mu \to 1-F$. On the other hand the Lorentz model fails as $\omega \to 0$ leading to an *incorrect* relative permeability of $1+F$. Indeed, the electromotive force driving the current through the ring and producing the magnetic response of the ring tends to zero as $\omega \to 0$. Besides that, at low frequencies the capacitive gap of the split-ring resonator prevents any current from flowing and, hence, there can be no magnetic response from the SRR array. This means that the limit $\mu \to 1$ as $\omega \to 0$ is the correct one in full accordance with (1). Thus the asymptotic behavior away from the resonance does not make the model described in (1) incorrect and does not make the Lorentz model correct. Neither (3) nor (1) alone cover the entire range of frequencies from zero to infinity.

The electromagnetic energy density for a Lorentz-type of media (3) has been derived earlier [5-8]. The corresponding expression, stemming from the low-frequency model (1) derived here is a complementary one, since its validity covers both the resonance region and the low-frequency region.

The magnetic induction $B(r,t)$ and the electric displacement $D(r,t)$ vectors can be introduced through the following constitutive relationships in the time-domain

$$B(r,t) = \mu_0 H(r,t) + M(r,t) \tag{4}$$

and

$$D(r,t) = \varepsilon_0 E(r,t) + P(r,t), \tag{5}$$

where $r$ is a spatial vector and $t$ is time. In (4) and (5) $\mu_0$, $\varepsilon_0$, $H$, $E$, $M$ and $P$ are the free-space permeability and permittivity, magnetic and electric field vectors, magnetization and polarization, respectively. Note that in (4) the magnetization is introduced in a slightly different way, compared to the standard definition $B(t) = \mu_0(H(t) + M(t))$ [11]. From Eqs. (1), (2), (4) and (5) the "equation of motion" for the magnetization is

$$\frac{\partial^2 M}{\partial t^2} + \gamma \frac{\partial M}{\partial t} + \omega_0^2 M = -\mu_0 F \frac{\partial^2 H}{\partial t^2}, \tag{6}$$

and for the polarization it is

$$\frac{\partial^2 P}{\partial t^2} + \nu \frac{\partial P}{\partial t} = \varepsilon_0 \omega_p^2 E. \tag{7}$$

Poynting's theorem [11] implies that

$$\operatorname{div}(E \times H) = -\frac{\partial}{\partial t}\left(\frac{\mu_0 H^2}{2} + \frac{\varepsilon_0 E^2}{2}\right) - H \cdot \frac{\partial M}{\partial t} - E \cdot \frac{\partial P}{\partial t}. \tag{8}$$

The material properties enter electromagnetic field energy density through the last two terms.

Hence, at this stage a number of paths can be taken. The most recent one involves an equivalent circuit (EC) approach [7] to calculating the energy. The latter yields a solution that does not demand any transformations and integrations of the type that will be done below. An approach involving equivalent circuits is not necessary, however, so Maxwell's equations and their consequences can be processed directly. By adopting this strategy, which will be referred to here as the electrodynamic (ED), the development will follow the path highlighted by Loudon. One of the points of interest is to see whether there is both qualitative and quantitative agreement between the two approaches.

Using the auxiliary field

$$\boldsymbol{C} = \boldsymbol{M} + \mu_0 F \boldsymbol{H} \tag{9}$$

Eq. (6) can be rewritten as

$$\omega_0^2 \mu_0 F \boldsymbol{H} \cdot \frac{\partial \boldsymbol{M}}{\partial t} = \frac{1}{2} \frac{\partial}{\partial t} \left\{ \left( \frac{\partial \boldsymbol{C}}{\partial t} \right)^2 + \omega_0^2 \boldsymbol{C}^2 - \omega_0^2 \mu_0^2 F^2 \boldsymbol{H}^2 \right\} + \\ + \gamma \left\{ \left( \frac{\partial \boldsymbol{C}}{\partial t} \right)^2 - \mu_0 F \frac{\partial \boldsymbol{H}}{\partial t} \cdot \frac{\partial \boldsymbol{M}}{\partial t} - \mu_0^2 F^2 \left( \frac{\partial \boldsymbol{H}}{\partial t} \right)^2 \right\}. \tag{10}$$

Using (7) now leads to [5, 6]

$$\omega_p^2 \varepsilon_0 \boldsymbol{E} \cdot \frac{\partial \boldsymbol{P}}{\partial t} = \left( \nu + \frac{1}{2} \frac{\partial}{\partial t} \right) \left( \frac{\partial \boldsymbol{P}}{\partial t} \right)^2. \tag{11}$$

After introducing the electromagnetic field energy density

$$w = w_E + w_M, \tag{12}$$

where $w_E$ and $w_M$ are the energy densities associated with the electric and magnetic fields, respectively, the use of Eqs (8)-(11) leads to the energy conservation law

$$\mathrm{div}(\boldsymbol{E} \times \boldsymbol{H}) = -\frac{\partial w_E}{\partial t} - \frac{\partial w_M}{\partial t} - P_L, \tag{13}$$

where $P_L$ is the power-loss. In (13) the energy density of the electric field is [5-7]

$$w_E(t) = \frac{\varepsilon_0}{2} \boldsymbol{E}^2 + \frac{1}{\omega_p^2 \varepsilon_0}\left(\frac{\partial \boldsymbol{P}}{\partial t}\right)^2. \tag{14}$$

The new result is that the energy density of the magnetic field is

$$w_M(t) = \frac{\mu_0(1-F)\boldsymbol{H}^2}{2} + \\ + \frac{1}{2\omega_0^2 \mu_0 F}\left\{\left(\frac{\partial \boldsymbol{M}}{\partial t} + \mu_0 F \frac{\partial \boldsymbol{H}}{\partial t}\right)^2 + \omega_0^2(\boldsymbol{M}+\mu_0 F\boldsymbol{H})^2\right\}. \tag{15}$$

This shows that the magnetic part of the energy density is strictly positive. Finally, the power-loss term is

$$P_L = \frac{\nu}{\omega_p^2 \varepsilon_0}\left(\frac{\partial \boldsymbol{P}}{\partial t}\right)^2 + \frac{\gamma}{\omega_0^2 \mu_0 F}\left(\frac{\partial \boldsymbol{M}}{\partial t} + \mu_0 F \frac{\partial \boldsymbol{H}}{\partial t}\right)\cdot\frac{\partial \boldsymbol{M}}{\partial t}. \tag{16}$$

## 3. Time-harmonic electromagnetic field

Equations (14), (15) and (16) are considerably simplified by adopting a time-harmonic electromagnetic field. This step introduces complex amplitudes through the definition

$$A(\boldsymbol{r},t) = \frac{1}{2}\left(\tilde{A}(\boldsymbol{r},\omega)\exp(-i\omega t) + c.c.\right), \tag{17}$$

where, $\omega$ is an angular frequency, $A$ stands for each of the quantities $\boldsymbol{H}$, $\boldsymbol{E}$, $\boldsymbol{M}$ and $\boldsymbol{P}$, and the phasor $\tilde{A}$ is the corresponding *complex* amplitude. The time-averaged electric and magnetic energy densities that follow directly from (14) and (15) are

$$\langle w_E \rangle = \frac{\varepsilon_0}{4}\left(1 + \frac{\omega_p^2}{\omega^2 + \nu^2}\right)|\tilde{\boldsymbol{E}}|^2 \tag{18}$$

and

$$\langle w_M \rangle = \frac{\mu_0}{4}\left(1 + F\frac{\omega^2\left[\omega_0^2\left(3\omega_0^2 - \omega^2\right) + \omega^2\gamma^2\right]}{\omega_0^2\left[\left(\omega_0^2 - \omega^2\right)^2 + \omega^2\gamma^2\right]}\right)|\tilde{H}|^2, \qquad (19)$$

respectively, where $\langle . \rangle$ denotes a time-average. Equation (18) is the time-averaged electric component of the energy. The latter is just a special case of the Lorentz-type of dielectric [5-8]. Equation (19) quantifies the ability of a SRR array, with a permeability function given by Eq. (1), to store magnetic energy. Thus it provides a measure of the response of the array. Note that $\langle w_E \rangle$ and $\langle w_M \rangle$ are strictly positive at all frequencies, regardless of the values of $\varepsilon(\omega)$ and $\mu(\omega)$ at the operating frequency $\omega$. This conclusion has been derived from the ED approach but it can also be drawn from the EC approach [7]. Nevertheless, as will be shown below, the ED approach not only has this property but it leads to the consistent limit in a dispersive lossless material.

If the losses are negligible, the time-averaged energy density, for a quasi-monochromatic (narrow-band) electromagnetic field, is [11]

$$\langle w \rangle = \frac{\varepsilon_0}{4}\frac{\partial(\omega\varepsilon(\omega))}{\partial\omega}|\tilde{E}|^2 + \frac{\mu_0}{4}\frac{\partial(\omega\mu(\omega))}{\partial\omega}|\tilde{H}|^2. \qquad (20)$$

Setting $\gamma = 0$ in (1) and $\nu = 0$ in (2), and using (20), leads to

$$\langle w \rangle = \frac{\varepsilon_0}{4}\left(1 + \frac{\omega_p^2}{\omega^2}\right)|\tilde{E}|^2 + \frac{\mu_0}{4}\left(1 + F\frac{\omega^2\left(3\omega_0^2 - \omega^2\right)}{\left(\omega_0^2 - \omega^2\right)^2}\right)|\tilde{H}|^2. \qquad (21)$$

This is *exactly* the same result as that obtained by setting $\nu = 0$ in (18) and $\gamma = 0$ in (19), which shows that, when the losses are negligible, the result derived here is consistent with the general formula for $\langle w \rangle$.

For a SRR array with a permeability given by (1) the EC approach [7] gives the magnetic component of the time-averaged energy density in the form

$$\langle w_M \rangle = \frac{\mu_0}{4} \mu_{eff} \left| \tilde{\boldsymbol{H}} \right|^2, \tag{22}$$

where $\mu_{eff}$ is the effective energy coefficient

$$\mu_{eff} = 1 + F \frac{\omega^2 (\omega_0^2 + \omega^2)}{(\omega_0^2 - \omega^2)^2 + \omega^2 \gamma^2}. \tag{23}$$

On the other hand the ED approach gives the following

$$\mu_{eff} = 1 + F \frac{\omega^2 \left[ \omega_0^2 (3\omega_0^2 - \omega^2) + \omega^2 \gamma^2 \right]}{\omega_0^2 \left[ (\omega_0^2 - \omega^2)^2 + \omega^2 \gamma^2 \right]}. \tag{24}$$

For a material with Lorentz-type of permeability, given by (3), the effective energy coefficient is [5-8]

$$\mu_{eff} = 1 + F \frac{\omega_0^2 (\omega_0^2 + \omega^2)}{(\omega_0^2 - \omega^2)^2 + \omega^2 \gamma^2}. \tag{25}$$

The time-averaged power $\langle P_L(t) \rangle$ absorbed by the SRR array, per unit volume, is

$$\langle P_L(t) \rangle = \frac{\mu_0}{4} \sigma_{eff} \gamma \left| \tilde{\boldsymbol{H}} \right|^2, \tag{26}$$

where $\sigma_{eff}$ is the effective energy loss coefficient. Using (16) (obtained from the ED approach to (1)) gives

$$\sigma_{eff} = \frac{F \omega^4}{(\omega_0^2 - \omega^2)^2 + \omega^2 \gamma^2}. \tag{27}$$

The same quantity, obtained from the Lorentz model (3) is

$$\sigma_{eff} = \frac{F \omega^2 \omega_0^2}{(\omega_0^2 - \omega^2)^2 + \omega^2 \gamma^2} \tag{28}$$

Eqs. (23), (24), (25), (27) and (28) are plotted in Fig. 1. As Fig. 1 (a) and (b) show, the effective energy coefficients (24) (obtained from the ED-approach to the model (1)) and the expression (25), originating from the Lorentz permeability model

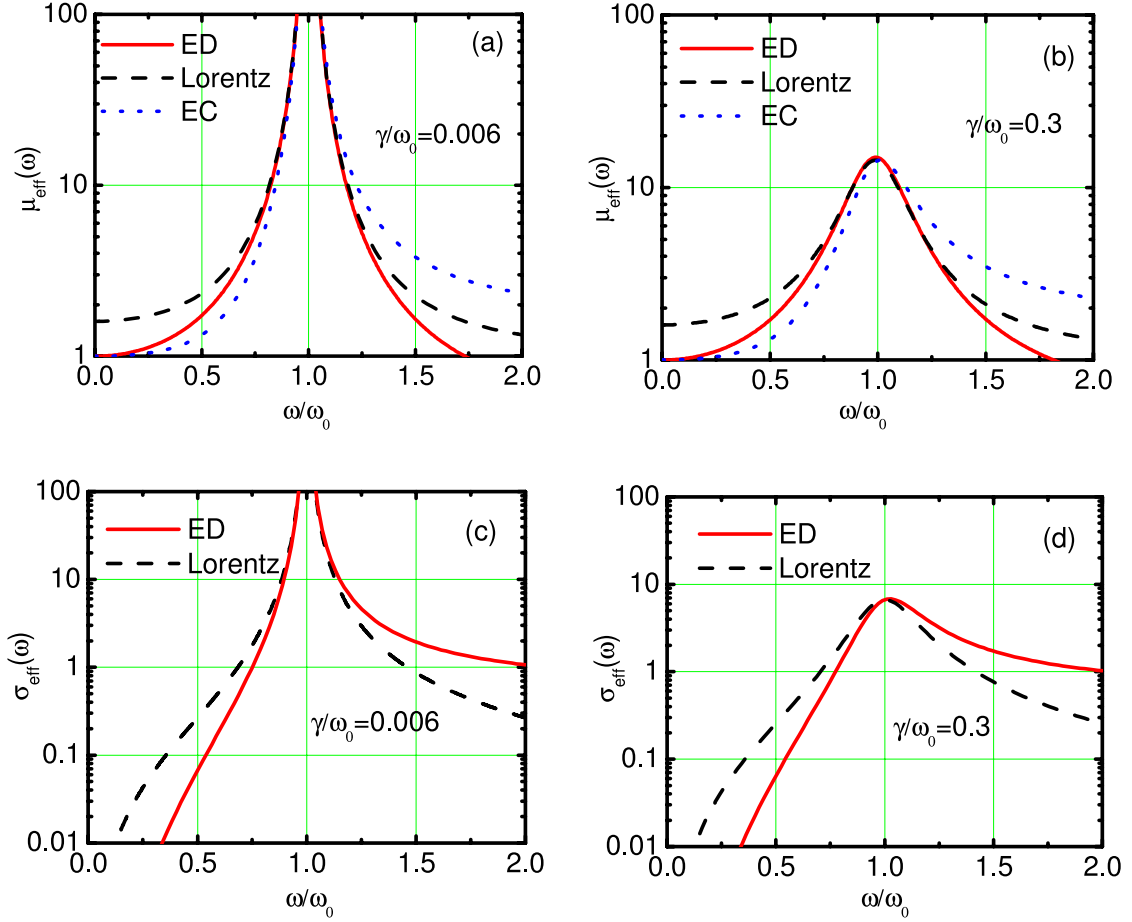

**Fig. 1 (Color online) Energy coefficient $\mu_{eff}(\omega)$ as given by Eqs. (23), (24) and (25) for (a) $\gamma/\omega_0 = 0.006$ and (b) $\gamma/\omega_0 = 0.3$. The three curves are labeled "EC", "ED" and "Lorentz" respectively. The value of the parameter $F$ is $F = 0.6$. Power loss coefficient $\sigma_{eff}(\omega)$ as given by Eqs. (27) ("ED") and (28) ("Lorentz") for (c) $\gamma/\omega_0 = 0.006$ and (d) $\gamma/\omega_0 = 0.3$.**

Eq. (3), are in agreement near the resonance $\omega \approx \omega_0$, as it should be expected. The difference between the two becomes evident away from the resonance region. As it has been already pointed out in the low frequency region $\omega < \omega_0$ preference should be given to (24) since the Lorentz model (3) does not have the correct low-frequency limit. On the other hand in the high-frequency region, $\omega > \omega_0$, the Lorentz permeability model, and its consequence Eq. (25) are expected to provide an adequate description. It has been pointed out [7] that the upper frequency limit above which the model (1) is no longer valid is the frequency at which the effective energy coefficient

(24) becomes smaller then one. Neglecting the losses, (24) yields $\omega < \sqrt{3}\omega_0$. The result obtained from the EC approach, (23) indeed gives $\mu_{eff} = 1$ at $\omega = 0$, as (23) and Fig. 1 show. Note, however, that even in the in the resonance region, where (24) and (25) are in agreement with each other, the difference between (24) and (25), on one hand, and (23), on the other, is significant as can be seen by examining the logarithmic scale. As Fig. 1 (c) and (d) show the *relative* difference between the loss coefficients (27) and (28) (resulting from (1) and (3), respectively) is significant away from the resonance $\omega = \omega_0$.

As pointed out earlier, in a dispersive, lossless material Eq. (23) is not compatible with the magnetic part of (21) [7]. This has been attributed to the fact that (1) is valid in the quasistatic limit only. But, as shown here, the expression for the magnetic energy density (19) and the magnetic part of (21) are in *perfect* agreement in a dispersive, lossless material. It can be concluded, therefore, that (19), (or, equivalently, (24)), obtained with the ED-approach, are more internally consistent.

## 4. Numerical results

The validity of (14) and (15) can be checked by considering the system shown in Fig. 2. It consists of a cylindrical Pendry-lens [3] and a wire dipole antenna. It has been shown recently [21] that a pair of dipole antennae, coupled by a "perfect" lens form an electromagnetic system that has a number of interesting properties.

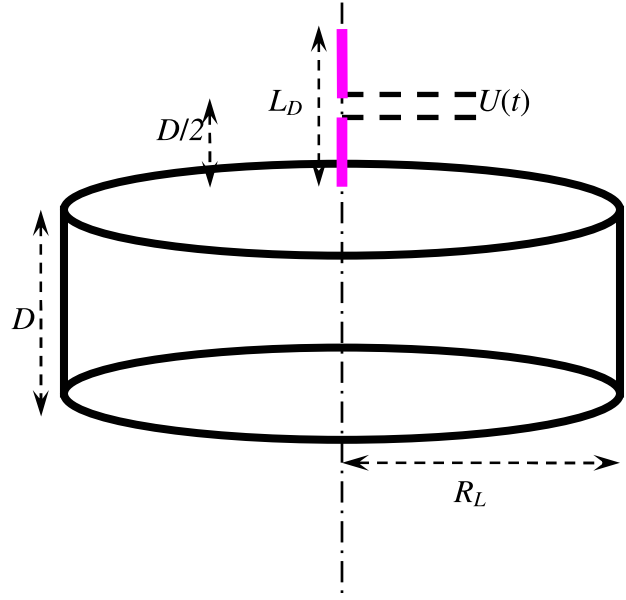

**Fig. 2 (Color online) A disc of thickness $D$ and radius $R_L$, made of a dispersive and lossy LHM, irradiated by a center-fed wire dipole antenna, of length $L_D$. The distance between the center of the antenna and the surface of the disc is $D/2$. The voltage feeding the dipole is $U(t)$.**

Fig. 2 shows an azimuthally symmetric arrangement consisting of a disc made of a left-handed metamaterial placed near a wire dipole antenna. This type of source is very convenient because it allows a direct connection between the voltage applied at the antenna terminals and the energy stored in the disc to be established easily. Assuming that the voltage is switched on at the time $t=0$, integrating (13) over the volume $V$ of the disc and over the time interval $[0, t]$ the energy conservation law becomes

$$W_{IN} = W + W_L, \qquad (29)$$

where

$$W_{IN} = -\int_0^t \oint_S [\mathbf{E} \times \mathbf{H} \cdot d\mathbf{S}]\, dt' \qquad (30)$$

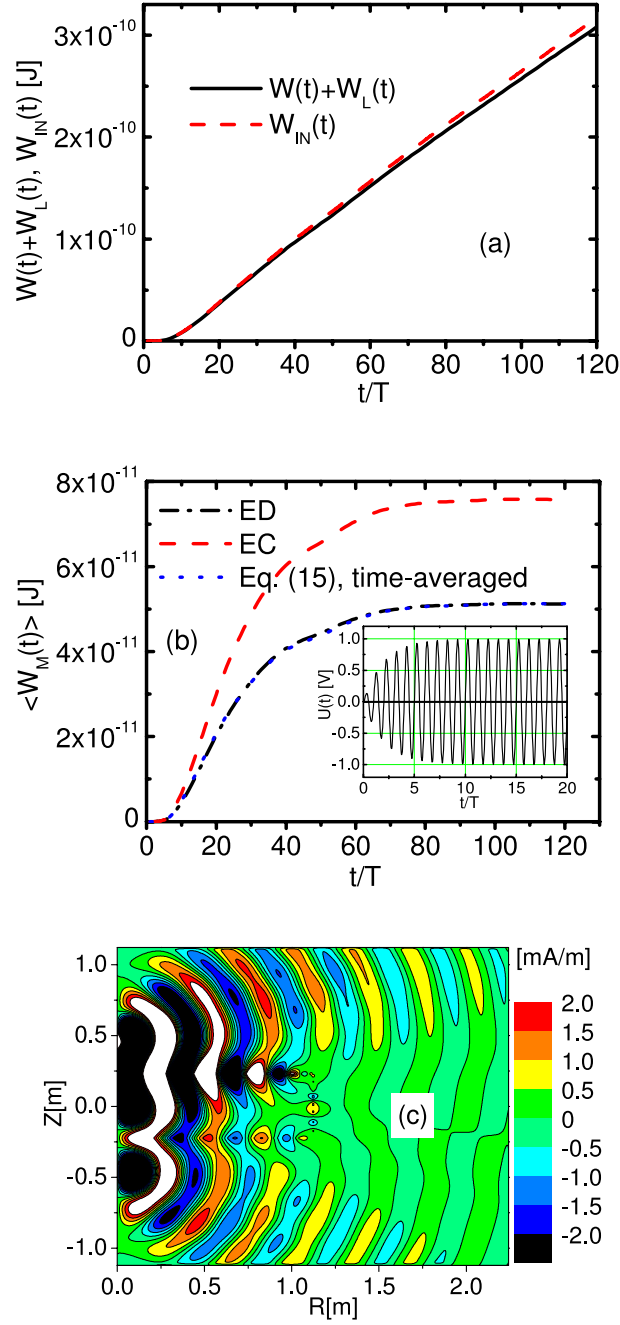

Fig. 3. (Color online) (a) the left- and the right-hand side of the energy conservation law (Eq. (29)) for the lensing arrangement from Fig. (2). The antenna feeding voltage is given by (33) (b) magnetic part of the stored energy obtained by integrating (22) over the volume of the lens with $\mu_{eff}$ given by Eqs. (23) ("EC") and (24) ("ED"). The same result obtained by time-averaging and subsequently integrating (15) over the volume of the disc is also shown ("Eq. (15), time aveargaed"). (c) magnetic field distribution $H_\varphi(R,Z)$ [mA/m] at the end of the computer run $t = 120T$. The image of the antenna is easy to see. $D = 44.85\,cm$, $R_L = 1.12\,m$, $L_D = 13.45\,cm$, $\omega/2\pi = 1\,GHz$, $\gamma/2\pi = 5\,MHz$, $\nu = 0$, $\omega_0/2\pi = 836.709\,MHz$, $\omega_p = \sqrt{2}\omega$, $F = 0.6$.

is the energy input to the disc ($S$ is the surface of the disc),

$$W = \int_V (w_E(t) + w_M(t)) dV \tag{31}$$

is the electromagnetic energy stored in the disc and

$$W_L = \int_0^t \int_V P_L(t') dV dt' \tag{32}$$

is the energy loss. The computational results given here use the FDTD method [22] and a thin-wire model [23] of the wire dipole, together with a feeding voltage in the form

$$U(t) = \left[1 - \exp\left(-\frac{\omega t}{4\pi}\right)\right] \sin(\omega t). \tag{33}$$

This corresponds to a sinusoidal waveform, of angular frequency $\omega = 2\pi/T$, being slowly switched on.

The set of equations, solved with the FDTD method, is

$$\varepsilon_0 \frac{\partial \bm{E}}{\partial t} = \nabla \times \bm{H} - \bm{J}, \tag{34}$$

$$\frac{\partial \bm{J}}{\partial t} + \nu \bm{J} = \omega_p^2 \varepsilon_0 \bm{E}, \tag{35}$$

$$\mu_0 \frac{\partial \bm{H}}{\partial t} = -\frac{1}{1-F}(\nabla \times \bm{E} + \bm{K}), \tag{36}$$

$$\frac{\partial \bm{K}}{\partial t} + \omega_0^2 \bm{M} = -\frac{\gamma}{1-F}(F \nabla \times \bm{E} + \bm{K}), \tag{37}$$

and

$$\frac{\partial \bm{M}}{\partial t} = \frac{1}{1-F}(\bm{K} + F \nabla \times \bm{E}). \tag{38}$$

Equation (35) results from (7) by introducing the electric current density $\bm{J} = \frac{\partial \bm{P}}{\partial t}$ in the latter. Equation (37) is obtained from (6) where the effective "magnetic current density" $\boldsymbol{K} = \frac{\partial \boldsymbol{C}}{\partial t}$ has been introduced. Equations (34) and (36) are the Maxwell's curl-equations. Equation (38) is obtained from the definition of the parameter $\boldsymbol{C}$ given by (8) and the subsequent use of (36). The set (34)-(38) gives the complete time-domain description of an electromagnetic field propagating in a dispersive and lossy LHM with the permittivity and the permeability of the latter given by (2) and (1), respectively.

The results obtained from the solution of the set (34)-(38) are presented in Fig. 3. The selected values of the resonant frequency and the plasma frequency are $\omega_0/2\pi = 836.709\,MHz$ and $\omega_p/2\pi = 1.414\,GHz$. These ensure that at the operating frequency $\omega/2\pi = 1\,GHz$ the values of the permittivity and the permeability functions are $\varepsilon(\omega) = -1$ and $\mu(\omega) = -1 + 0.0556i$. As Fig 3 (a) shows, the energy conservation law (29), with the magnetic part of the stored energy calculated from Eq. (15) is satisfied to a high degree of accuracy, with the maximum relative error being less than 3%. Equation (19) is in an excellent agreement with the time-averaged version of (15), as can be seen from Fig. 3 (b). This is because the electromagnetic field is, in fact, monochromatic. Expression (23), resulting from the EC-approach in this case overestimates the magnetic part of the stored energy by 50%, as Fig. 3(b) shows. The dynamics of the stored energy, presented in Fig. 2, show that while the amplitude of the feeding voltage reaches a stationary state for about 10 periods of the carrier frequency, the duration of the relaxation period for the stored energy is more than 60 periods. This feature is related to the finite size of the lens and depends on the losses [24]. Therefore, the formation of a stationary electromagnetic field distribution inside the lens and, consequently, the formation of the image (Fig. 3b) can be regarded as a "slow" process.

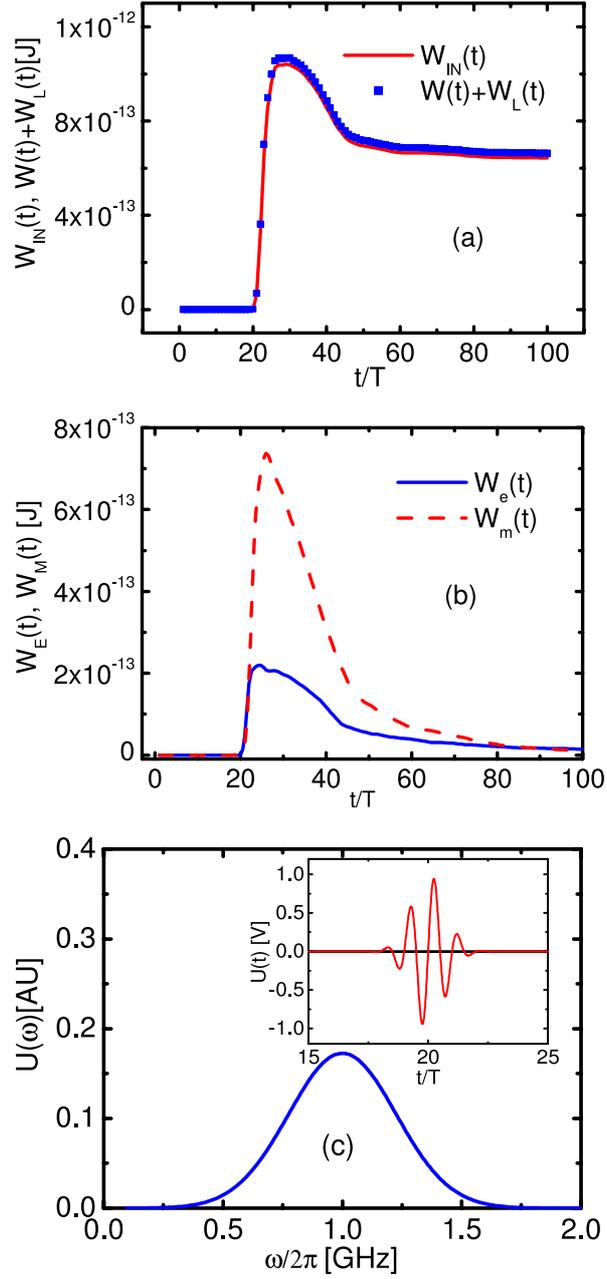

Fig. 4. (Color online) (a) the left- and the right-hand side of the energy conservation law (29) with the antenna feeding voltage given by (39). (b) dynamics of the stored electric and magnetic components of the energy obtained by integrating (14) and (15) over the volume of the disc. (c) the spectrum of the feeding voltage. The inset shows the feeding voltage waveform. All the system parameter values are as in Fig. 3.

Fig. 4(a) shows the energy conservation law for a short pulse form of feeding voltage

$$U(t) = \exp\left[-(t/T - 20)^2\right] \sin[2\pi(t/T - 20)] \tag{39}$$

centered at $t = 20T$. The overlap between the pulse spectrum Fig 4(c) and resonant curve shown in Fig. 1(a) is strong in this case. As in Fig. 3(a), the energy conservation law is satisfied again to a high degree of accuracy. The corresponding electric and magnetic energies are shown in Fig. 4(b). The magnitude of the magnetic stored energy is larger than the magnitude of the electric energy, which is consistent with the resonant structure of $\mu(\omega)$.

## 5. Conclusions

Expressions for the energy density and energy losses in a dispersive and lossy left-handed metamaterial, consisting of an array of wires and an array of SRRs are derived. An electromagnetic field with arbitrary time-dependence is considered. Under conditions of negligible losses, the result for the magnetic part of the energy (the energy stored in the SRR array) obtained here is in full agreement with the general formula, valid for a lossless dispersive material. In the resonance region, the new result for the magnetic part of the stored electromagnetic energy is shown to map quantitatively onto the result obtained from the Lorentz permeability model. The power-loss terms, resulting from the two permeability models considered, however, quickly diverge away from the resonance. Exact FDTD-solutions of Maxwell's equations show that the energy conservation law is satisfied to a high degree of accuracy, thus validating the analytical results obtained.

## Acknowledgements

This work is supported by the Engineering and Physical Sciences Research Council (UK) under the Adventure Fund Programme.